# Design and analysis of annulus core few mode EDFA for modal gain equalization

Ankita Gaur, *Member OSA,* Vipul Rastogi, Senior *Member OSA*

*Abstract*—Few-mode fiber amplifier is widely under study to overcome the issue of internet traffic in optical communication. This article proposes annulus core few-mode erbium doped fiber (FM-EDF) with annulus or extra annulus doping for amplification of the $LP_{01}$, $LP_{11}$, $LP_{21}$, and $LP_{31}$ signal mode groups with low differential modal gain (DMG). Our simulations confirm that extra annulus doping helps in reducing DMG of higher order mode groups. We have achieved less than 2.2 dB DMG over C-band for 4-mode groups using extra annulus doping. The proposed EDF would be useful for space division multiplexing (SDM) based optical fiber communication system.

## I. Introduction

STATISTICS of capacity with time suggest that increase in internet traffic may soon lead to capacity crunch as we are approaching to capacity limits of the single mode optical fiber systems. SDM is an effective way of increasing transmission capacity of optical fiber communication by increasing the number of channels [1]. There are two ways to incorporate SDM in optical communication system : multi-core fiber and few-mode fiber. Recently 112 Tb/s transmission capacity using a seven-core fiber [2] and 57.6 Tb/s net capacity using few mode fiber [3] have been reported. In addition to this, the development of SDM requires SDM integration based transmitter, receiver and optical amplifier. The most commonly used optical amplifier for optical communication is erbium doped fiber amplifier (EDFA). An aggregate capacity of 1.2 Tbit/s over a 250-GHz bandwidth using a three-core microstructured fiber with inline amplification via parallel single-mode EDFAs was demonstrated in [4] with maximum transmission distance of 4200 km. For cost reduction, SDM requires simultaneous amplification of the individual guided modes within the amplifier.

To enable few-mode fiber based SDM optical communication system, it is necessary to develop FM-EDFA which could provide high gain, low DMG and mode number scaling. Control over DMG mainly depends on erbium-doping profile and pump intensity distribution [5]. A system of multimode pump configuration has been demonstrated for controlled mode dependent gain at 1530 nm with DMG less than ±0.5 dB for two mode groups using uniformly-doped FM-EDF [6]. Several FM-EDFs using various controlled erbium doping profiles have been proposed. The ring-doped step-index fibers have been studied to minimize the DMG of the signal mode groups [7], [8]. The extra-annulus doping has also been studied for step-index fiber with or without trench and for graded-index fiber with or without trench to minimize DMG [9]. The improvement of DMG in FM-EDFA by employing graded-index ring-core erbium-doped fiber instead of conventional step-index fiber has been studied [10], [11].

Recently, a ring core FMF which supports three mode groups $LP_{01}$, $LP_{11}$, $LP_{21}$ and does not support $LP_{02}$ and $LP_{22}$ mode groups (to avoid mode conversion between $LP_{01}$ and $LP_{02}$ at discontinuity point) has been studied to minimize the differential mode delay (DMD) [12]. In this paper, we have studied an annulus core EDFA which supports $LP_{01}$, $LP_{11}$, $LP_{21}$ and $LP_{31}$ mode groups only. The proposed EDF is studied with annulus and extra-annulus doping. The design parameters dependence of effective index ($n_{eff}$) has been studied. The $LP_{01}$ pump mode has good overlaps with the ring-shaped signal modes of FM-EDF. We have introduced extra-annulus doping in annulus core fiber and have shown that extra annulus doping helps in reducing DMG. In this work, we have scaled up the signal mode groups of annulus core EDFA to 4 while maintaining effective indices separation between adjacent mode groups ($\Delta n_{eff}$) greater than $5.5 \times 10^{-4}$. We have achieved less than 2.2 dB DMG over C-band using extra annulus doping. The proposed fiber could be useful in FMF based optical communication system.

## II. Fiber Design

The proposed FM-EDF structure is schematically shown in Fig. 1. It consists of two core segments: one is uniform low index core ($r < a$) and other is annulus high index core ($a < r < b$). The central low-index core segment and the cladding are made up of silica of refractive index $n_{silica}$. The annulus core segment of width $d$ is made up of Ge doped silica. The relative refractive index difference between the high-index ($n_1$) core and silica ($n_{silica}$) is defined as:

$$\Delta = \frac{(n_1^2 - n_{silica}^2)}{2n_1^2}$$

Vipul Rastogi is with the Department of Physics, Indian Institute of Technology Roorkee, Roorkee, India (email: vipul.rastogi@osamember.org).

Ankita Gaur is with the Department of Physics, Indian Institute of Technology Roorkee, Roorkee, India (email: ankitagaur.phy@gmail.com).

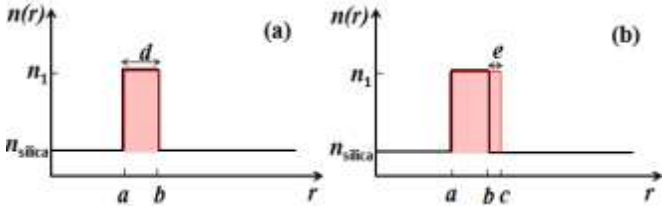

Fig. 1. Schematic of refractive index profile of the proposed EDF (shaded portion represents doping region) (a) annulus doping (b) extra annulus doping

The proposed design has been studied with annulus and extra annulus $Er^{+3}$ doping as shown in Fig. 1. We have chosen $LP_{01}$ mode at 980 nm wavelength as pump mode. The standard techniques of phase mask or spatial light modulator could be used for pump excitation. The mode profiles of different signal mode groups and $LP_{01}$ pump mode of FM-EDF with $\Delta = 1\%$, $a = 5.2$ μm, $d = 2.7$ μm, $e = 0.5$ μm are shown in Fig. 2. Mode effective indices and corresponding mode profiles of the fiber have been calculated by using Transfer Matrix Method [13]. The power confinement in $Er^{+3}$ doped region and nearly equal overlap of the chosen pump mode with different signal mode groups enable to achieve high gain and low DMG. We can see that $LP_{01}$ pump mode will have equal overlaps with even and odd modes of the same mode group. Our calculations also show that it has similar overlaps with different signal mode groups.

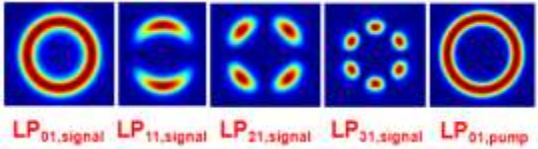

Fig. 2. Mode profiles of different signal mode groups and pump mode of FM-EDF

### III. NUMERICAL SIMULATION AND RESULT

The gains of FM-EDFA have been calculated by using mathematical modeling given in Ref. [6]. For the calculation of gain, the input signal power in each orientation and polarization of signal mode groups has been chosen as 30 μW and erbium ion concentration used is $N_0 = 1\times10^{25}$ m$^{-3}$. The absorption and emission cross-sections used at 1530 nm are $4.68\times10^{-25}$ m$^2$ and $5.54\times10^{-25}$ m$^2$ respectively and the pump absorption cross-section used at 980 nm is $1.879\times10^{-25}$ m$^2$ [14]. The simulations have been performed for annulus and extra annulus doping.

Recently, a ring core FMF has been studied for three mode groups to reduce DMD for SDM communication system [12]. To incorporate such fiber in SDM optical system, the design of corresponding ring core FM-EDFA is required. We have designed 3-mode group ring core FM-EDFA for such FMF. Fig. 3 (a) shows the variation of gains of $LP_{01}$, $LP_{11}$, and $LP_{21}$ mode groups of FM-EDF with its length. The parameters used are: $\Delta=1\%$, $a = 3.5$ μm, $d = 2.7$ μm and $e = 0.5$ μm. The modes of the fiber corresponding to these parameters have sufficient mode spacing ($> 1.1\times10^{-3}$) in order to avoid mode coupling due to macro-bending [15]. For annulus doping, overlaps of pump with $LP_{01}$, $LP_{11}$ and $LP_{21}$ signal mode groups of FM-EDF at 1530 nm wavelength are $8.38\times10^9$, $8.53\times10^9$ and $8.19\times10^9$ respectively and for extra annulus doping,

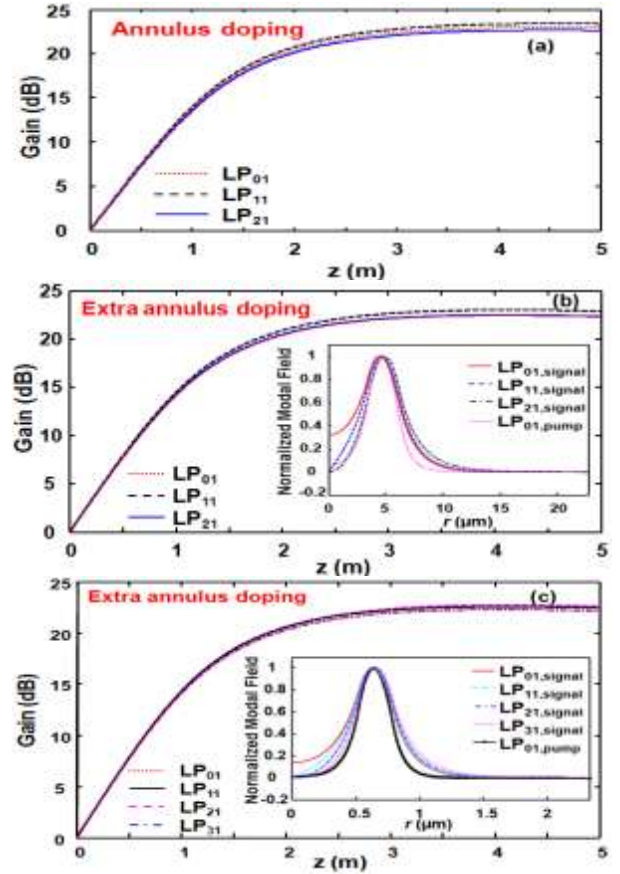

Fig. 3. (a) Variations of gains of $LP_{01}$, $LP_{11}$ and $LP_{21}$ mode groups vs the EDF length with annulus doping (b) Variation of gains of $LP_{01}$, $LP_{11}$ and $LP_{21}$ mode groups vs the EDF length with extra annulus doping (mode field of $LP_{01,signal}$, $LP_{11,signal}$, $LP_{21,signal}$ signal mode groups and $LP_{01,pump}$ pump mode are shown in inset) (c) Variation of gains of $LP_{01}$, $LP_{11}$, $LP_{21}$ and $LP_{31}$ mode groups vs the EDF length with extra annulus doping (mode field of $LP_{01,signal}$, $LP_{11,signal}$, $LP_{21,signal}$, $LP_{31,signal}$ and $LP_{01,pump}$ are shown in inset)

overlaps of pump with $LP_{01}$, $LP_{11}$ and $LP_{21}$ signal mode groups of FM-EDF at 1530 nm wavelength are $8.51\times10^9$, $8.68\times10^9$ and $8.38\times10^9$ respectively. The result shows that more than 20 dB gain for each mode group and less than 0.8 dB DMG have been achieved using input pump power of 150 mW and amplifier of length longer than 2 m with annulus doping. This is due to almost equal overlaps of pump with signal mode groups. Fig. 3 (b) shows that using extra annulus doping, DMG of 3 mode groups becomes less than 0.6 dB. This indicates that DMG could be further reduced using extra annulus doping. This happens because signal mode profiles of higher-order mode groups extend more towards extra annulus region as shown in the inset of Fig. 3 (b), so extra annulus doping enhances the gains of higher-order mode groups in comparison to those of lower-order mode groups, which results in decrease of DMG. The work could be extended to even higher mode groups. Fig 3 (c) shows the gain of 4 mode groups $LP_{01}$, $LP_{11}$, $LP_{21}$ and $LP_{31}$ with FM-EDFA length using extra annulus doping and input pump power of 200 mW. The fiber parameters used are $\Delta = 1\%$, $a = 5.2$ μm, $d = 2.7$ μm, $e = 0.5$ μm. The overlaps of pump with $LP_{01}$, $LP_{11}$, $LP_{21}$ and $LP_{31}$ signal mode groups of FM-EDF at 1530 nm wavelength using extra annulus doping are $6.25\times10^9$, $6.32\times10^9$, $6.33\times10^9$ and $6.16\times10^9$ respectively. The results show more than 20 dB gain with less than 0.5 dB DMG for fiber length greater than 2 m. The results and parameters corresponding to Figs. 3 (b) and (c) suggest that the number of supported modes and DMG control depends on various parameters of the EDF. So, the parameters of amplifier should be optimized for minimum DMG while maintaining sufficient mode spacing $\Delta n_{\text{eff}}$ between adjacent mode groups. Therefore, we have studied the effect of different parameters on DMG of signal mode groups.

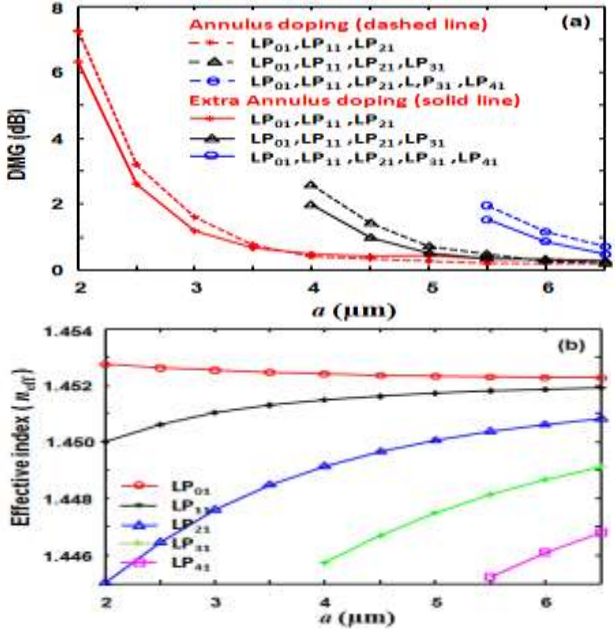

Fig. 4. (a) Variation of DMG with central core radius $a$ using input pump power 250 mW and fiber length 3 m at 1530 nm signal wavelength. (b) Variation of $n_{\text{eff}}$ with $a$ parameter at 1530 nm.

The variation of DMG with central core radius $a$ is shown in Fig. 4 (a). The variation of mode effective indices with $a$ is shown in Fig. 4 (b) at 1530 nm wavelength. The other parameters used in calculation are: $\Delta = 1\%$, $d = 2.7$ μm, $e = 0.5$ μm. The results show that DMG and mode spacing $\Delta n_{\text{eff}}$ decrease with increase in $a$. For $a = 3.5$ μm, 3-signal mode groups are supported and their DMG is 0.76 dB with annulus doping and 0.66 dB with extra annulus doping while the value of $\Delta n_{\text{eff}}$ is more than $1.1\times10^{-3}$. For $a = 5$ μm, there are 4-mode groups and their DMG is 0.72 dB with annulus doping and 0.51 dB with extra annulus doping while effective index spacing is greater than $6\times10^{-4}$. For $a = 6.5$ μm, DMG of five mode groups is 0.72 dB with annulus doping and 0.49 dB with extra annulus doping and effective index spacing is greater than $3.5\times10^{-4}$. The results also show that there is a decrease in DMG with extra annulus doping in comparison to annulus doping.

Figs. 5 (a) and (b) show the effect of $\Delta$ on DMG and mode effective indices. The other parameters used in calculation are: $a = 5.2$ μm, $d = 2.7$ μm, $e = 0.5$ μm. The results show that DMG decreases with increase in $\Delta$ and there is no significant change in $\Delta n_{\text{eff}}$. The results also show the significant effect of extra annulus doping. For $\Delta = 0.9\%$ and extra annulus doping, DMG of 4 mode groups is 0.71 dB with $\Delta n_{\text{eff}}$ greater than $5.7\times10^{-4}$. Using $\Delta = 1.3$, the DMG of 4 mode groups can be brought down to 0.21 dB. However, a higher value of $\Delta$ would also result in propagation loss due to high Ge concentration. Thus, $\Delta = 1\%$ is optimum from point of view of DMG control and fiber manufacturing.

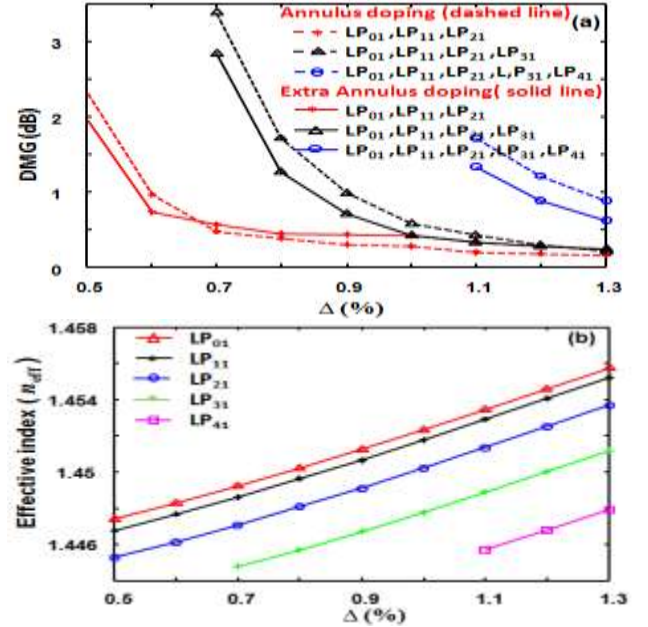

Fig. 5. (a) Variation of DMG of $LP_{01,FMF}$, $LP_{11,FMF}$ and $LP_{21,FMF}$ with $\Delta$ using input pump power 250 mW and fiber length 3 m at 1530 nm signal wavelength. (b) Variation of $n_{\text{eff}}$ with $\Delta$ at 1530 nm.

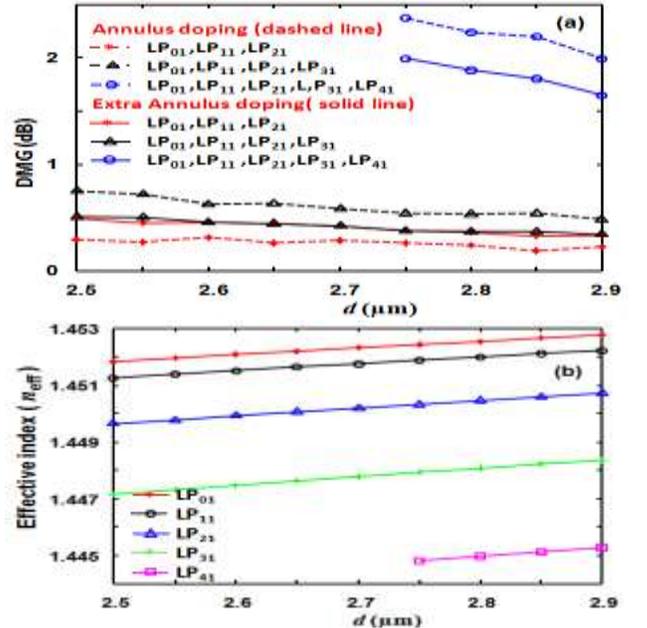

Fig. 6. (a) Variation of DMG with $d$ using input pump power 250 mW and fiber length 3 m at 1530 nm signal wavelength (b) Variation of $n_{\text{eff}}$ with $d$ at 1530 nm.

Figs. 6 (a) and (b) show the tolerance of annulus core width $d$ on DMG and mode indices. The other parameters used in calculation are: $\Delta = 1\%$, $a = 5.2$ μm, $e = 0.5$ μm. The results show that DMG decreases with increase in $d$ and $\Delta n_{\text{eff}}$ also decreases with increase in $d$. However, in the range 2.6 μm $< d <$ 2.7 μm, DMG of 4 mode groups varies from 0.46 dB to 0.42 dB with extra annulus doping and $\Delta n_{\text{eff}}$ is greater than $5.5\times10^{-4}$. It means, there is only a small variation of 0.04 dB in DMG of 4 mode groups with 0.1 μm change in $d$.

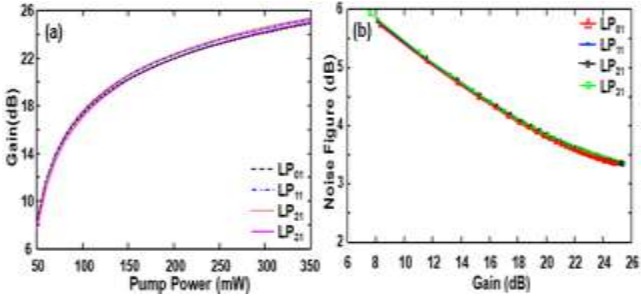

Fig. 7. (a) Variation of gains $LP_{01}$, $LP_{11}$, $LP_{21}$ and $LP_{31}$ mode groups with pump power at 1530 nm signal wavelength (b) Corresponding noise figure

The variation of pump power on gains of different signal mode groups is shown in Fig. 7 (a). The optimized parameters used in calculation are: $\Delta = 1\%$, $a = 5.2$ μm, $d = 2.7$ μm and $e = 0.5$ μm. The results show that for pump power $P_{pump} \geq 150$ mW, more than 20 dB gain for each mode group and less than 0.5 dB differential modal gain (DMG) with sufficient mode spacing $> 5.5\times10^{-4}$ to avoid mode coupling have been achieved using amplifier of length 3 m. DMG varies from 0.3 dB to 0.5 dB with pump power variation from 150 mW to 350 mW, which shows that any fluctuation in pump power does not significantly affect the DMG. Fig. 7 (b) shows the convergence of noise figures towards 3.3 dB with increase in modal gain.

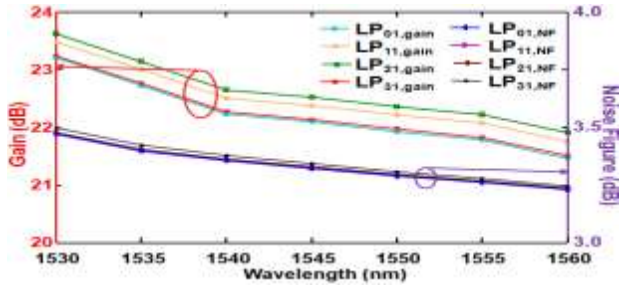

Fig. 8 Variation of gains and noise figures of $LP_{01}$, $LP_{11}$, $LP_{21}$ and $LP_{31}$ mode groups with wavelength

The study of modal gains and noise figures (NF) of $LP_{01}$, $LP_{11}$, $LP_{21}$ and $LP_{31}$ mode groups of FM-EDFA over C-band (7 equally spaced wavelengths with equal input signal power of 30 μW) at 250 mW input pump power and 3 m FM-EDF length is shown in Fig. 8. The study shows that DMG between the four mode groups corresponding to single wavelength is below 0.45 dB and the gain difference for a mode group over C-band is less than 1.75 dB. Thus, overall DMG is below 2.2 dB over entire C-band with $\Delta n_{eff} > 5.5\times10^{-4}$ and noise figure less than 3.5 dB. The gain difference can be further minimized by using gain equalization filter [16].

## IV. CONCLUSION

We have studied an annulus core FM-EDF with extra annulus doping for applications in SDM optical communication system. We have shown that DMG of 3 mode groups is less than 0.6 dB with effective indices separation greater than $1.1\times10^{-3}$. We have also investigated the effect of annular core width, relative index difference and central core radius on DMG and mode spacing using annulus and extra annulus doping. Extra annulus doping plays a significant role in minimizing DMG. The study shows that less than 2.2 dB DMG over C-band for 4 mode groups EDF could be achieved with mode spacing $> 5.5\times10^{-4}$. The proposed configuration would be useful for SDM system.